\documentclass{phostproc}

\title{Current problems in stellar evolution}
\author{Gael Buldgen$^{1,2}$}

\affiliation{$^{1}$ School of Physics and Astronomy, University of Birmingham, Edgbaston, Birmingham B15 2TT, UK.\\
$^{2}$ Observatoire de Genève, Université de Genève, 51 Ch. Des Maillettes, CH-1290 Sauverny, Suisse.}

\shorttitle{Current problems in stellar evolution}
\shortauthors{G. Buldgen}

\abs{The theory of stellar evolution plays a central role in astrophysics as stellar models are used to infer properties for Galactic and Extragalactic stellar populations as well as exoplanetary systems. However, despite decades of experience, stellar models still face major issues linked to transport processes of chemicals and angular momentum. This review will focus on some of the processes responsible for the most sizable uncertainties in stellar models such as for example convection, rotation and mass loss. The presentation will discuss their implementation, their impact on theoretical predictions and how various observational constraints can help us gain insight on the physics inside stars and face the current challenges of the theory of stellar evolution.}

\begin{document}

\maketitle

\section{Introduction}

Stars are key elements of the Universe, they play the role of distance and age indicators for other fields of astrophysics and are at the origin of the chemical evolution of the Universe. The theory of stellar evolution is one of the oldest fields of astrophysics and has enjoyed decades of confrontation between theory and observations, especially since the capability to compute numerically theoretical stellar models on a large scale. 

Despite tremendous progress on the observational side, the availability of very high quality interferometric, spectroscopic and seismic data for a large number of stars, and the monumental increase of computational power, key issues still remain open in the theory of stellar structure and evolution. 

In this review, I will briefly present some of the key uncertainties of stellar models, their impact on observations, how they can be studied and constrained and provide further references to more detailed studies. I focus on macroscopic aspects namely rotation, convection and radiatively driven mass loss in stellar winds of isolated stars.  

I start in section \ref{secModels} by discussing the various scales of stellar evolution and how they are linked to some of the most thorny issues of the current generation of stellar models. In section \ref{secConvection}, I will discuss the specific issues linked to the modelling of convective transport in stellar interiors. I then turn towards the modelling of stellar rotation and its implication for stellar structure and evolution before discussing the importance of mass loss. I then conclude on future prospects for constraining these uncertain processes and the subsequent advances in theoretical stellar modelling.

\section{Stellar models and their uncertainties}\label{secModels}

The goal of stellar models is to reproduce, as accurately as possible, the structure and evolution of stars throughout the various stages of their lives. This somewhat naive statement contains however a range of difficulties, due to the fact that stars are subject to physical phenomena with very different spatial and temporal scales. Despite the increase of computational power, it is clearly impossible to solve the full system of hydrodynamical equations to follow the evolution of a star. Consequently, ``classical'' quasi-static stellar models will evidently still be in use on foreseeable timescales. 

These models have the strong advantage to be very easy to compute on large scale while still being in ``reasonable'' agreement with observations. For example, the good agreement of standard solar models with helioseismic constraints proves that, at least in the very controlled experimental framework of the Sun, they perform very well and provided reliable and testable predictions for the solar neutrinos. This is however not the case for other regions of the HR diagram. 

The origin of these issues is the difficulties of static stellar models to properly reproduce the transport properties of real stars. Indeed, stars are interacting, rotating, tridimensional objects, while the quasi-static models assume spherical symmetry, are most often non-rotating and non-interacting objects. The models are governed by the following simple ordinary differential equations
\begin{eqnarray}
\frac{dr}{dm}=\frac{1}{4\pi r^{2}\rho}, \\
\frac{dP}{dm}=-\frac{Gm}{4\pi r^{4}}, \\
\frac{dL}{dm}= \epsilon_{\mathrm{nuc}} -\epsilon_{\nu} + \epsilon_{\mathrm{grav}}, \\
\frac{dT}{dm}=- \frac{GmT}{4\pi Pr^{4}}\nabla ,
\end{eqnarray}

where the mass coordinate, $m$, has been used as the independent variable and I have defined $\rho$, the local density, $r$ the radial coordinate for a given mass $m$, $P$ the local pressure, $L$ the local luminosity, $T$ the local temperature, the temperature gradient defined as $\nabla=\frac{d \ln T}{d \ln P}$, and $\epsilon_{\mathrm{nuc}}$, $\epsilon_{\mathrm{grav}}$  and $\epsilon_{\nu}$, the specific energy generation rate from nuclear reactions, gravitational contraction and the energy losses from neutrinos, respectively. These equations are solved at a given time, then, the evolution of the chemical abundances in each layer is considered from a set of additional equations which govern the abundance changes during an evolutionary timestep. Once the equilibrium abundances have been redetermined, the structure is reintegrated for the obtained chemical profiles. 

As a consequence of this separation of timescales, transport mechanisms are not included directly in the stellar structure equations. They must be first identified, then implemented as accurately as possible (in 1D) in the stellar evolution code. For that purpose, hydrodynamical simulations provide a good complementary approach, in the form of a snapshot (i.e. a very short duration and/or a very small part of the star) of what these processes may look like. It should be noted, however, that hydrodynamical simulations, although much more realistic than the approximate formalisms used in 1D models, are not often computed in the range of physical conditions found in stellar interiors. Moreover, the science of modelling turbulent flows with numerical simulations is also subject to its own evolution, like any research field and will thus influence the conclusions we draw for stellar models.  

A tedious issue with these processes is not only to identify them individually, but also to understand their interactions, coupling and dependences on some key physical ingredients of stellar models. For example, the development of hydrodynamical instabilities is affected by the presence of strong chemical composition gradients which are in turn affected by microscopic diffusion, itself affected by the values of radiative opacities which are also influenced by the equation of state of stellar material. Figure 1 of \citet{Mathis} shows an example of such coupling between micro and macro-physical processes and studies of this ``mic-mac'' connexion in the case of thermohaline convection in massive stars has been presented in \citet{Maeder2013, Deal2016, Vauclair2016}. 

In general, it is common to separate the uncertainties of stellar models in two classes: the macro-physical processes and the micro-physical ingredients. In what follows, I list some of these ingredients and provide additional references.

\subsection{Macro-physical processes}

Macrophysical processes include phenomena such as  convection, rotation, gravity waves, magnetic fields and mass loss. Their depiction in stellar models is particularly difficult since they are associated with the breaking of spherical symmetry and a wide range of spatial and temporal scales which cannot be fully modelled on evolutionary timescales.  

All these processes have been intensively investigated throughout the history of stellar evolution. Convection, for example, is an excellent example of the difficulties of stellar modellers to tackle hydrodynamical phenomena that I will further discuss below. I refer the interested reader to the following papers for additional informations \citep{Spruit1990,Xiong1997,Canuto1998, Brummell2002,Noels2010,Canuto2011,Canuto2011b,Noels2013,Kupka2017}. Besides convection, stellar modellers have also been very preoccupied with the modelling of rotation. As an example, the meridional circulation, which was theorized as a consequence of the thermal imbalance induced by the breaking of spherical symmetry due to rotation (Von Zeipel's theorem) was one of the most debated problem of the past century. As we will see below, the inability of the current stellar models to reproduce the abundance patterns in clusters and the need for an additional process to reproduce both the solar rotation profile and the seismic observations in post main-sequence stars advocated for an improper modelling of rotation in low-mass stars. In the solar case, two scenarios were proposed, related to either the effect of magnetic fields \citep{Eggenberger2005} or gravity waves \citep{Charbonnel1999}, while the issue remains open for post main-sequence stars as multiple scenarios could be at play in the subgiant phase and the red giant phase \citep{Spada2016, Pincon2017}. 

As discussed above, the coupling of these processes with additional phenomena adds another layer of complexity to their modelling in stellar evolution code. Such a phenomenon is mass loss, which is strongly influenced by other dynamical phenomena such as rotation and pulsations. The current issues with mass loss are related to the diversity of processes that fall under that denomination and the fact that they are most often included through approximate empirical formulas in stellar evolution code. Here again, hydrodynamical simulations of stellar winds and high quality observational constraints play a crucial role in providing more robust formulations. 

\subsection{Micro-physical ingredients}

Microphysical ingredients include aspects of the models such as the nuclear reaction rates, the opacity tables, the chemical abundances, the equation of state and the modelling of miscroscopic diffusion. Their uncertainties are not only linked to hydrodynamical questions but also to more fundamental physical aspects such as the depiction of the interaction of the plasma constituents when computing the equation of state, the effects of partial ionization, quantum corrections\citep{Schlattl2003} and radiative accelerations on diffusion coefficients \citep[see][ for a discussion]{Turcotte1998a, Turcotte1998b, Schlattl2002,Gorshkov2010,Deal2018}. 

A good example of these issues is the current debate in the opacity community which followed the disagreement between the experimental measurements of \citet{Bailey2015} and the theoretical computations of other groups \citep[see][for example]{Lynas2018, Pain2018,Pain2019}. These discussions are to be put in perspective with the solar modelling issues which followed the downward revision of the solar heavy element abundances by \citet{AGSS} and the inability of models including the current opacity tables to reproduce the observations of B stars exhibiting pulsations without additional modifications \citep{Daszy2005,Salmon2012, Iglesias2015,Daszy2017,Daszy2018,Hui2018}.

\section{Convection}\label{secConvection}

Under certain conditions, large scale motions can occur in stellar interiors. These motions are described as convective, namely, the properties of the fluid are transported by turbulent macroscopic motions of the fluid itself. This provides a very efficient mechanism for the homogenization of the chemical composition and the outward transport of energy in the deep layers. Convection occurs for every range of mass and chemical composition throughout the evolution of the star, either in the deep or outer layers. One then speaks of convective envelopes or convective cores. In advanced stages of nuclear burning, intermediate convective zones can also appear in shells undergoing nuclear fusion.

The main issue of modelling convective motions is that they are highly turbulent in the conditions of stellar interiors, far beyond what can be achieved in numerical simulations and experiments. Furthermore, the short timescales of stellar convection imply that the 1D stellar models will rely on approximate formalism to define averages for a very multi-dimensional, multi-scale physical process. The first issue encountered in stellar models is to define a proper criterion to be able to tell, from a purely local and static diagnostic of the physical properties, whether a layer of stellar material will be convective or not. This analysis can be made by using the local temperature and chemical composition gradients. Depending on the situation, various instabilities can occur. This is illustrated in figure 2 of \citet{Salaris}, which can be used as a visual summary of the various criteria implemented in stellar evolution codes.

The first criterion is the so-called Schwarzschild criterion, which states that if locally one has
\begin{equation}
\nabla_{\mathrm{rad}}>\nabla_{\mathrm{ad}},
\end{equation}
with $\nabla_{ad}=\frac{d \ln T}{d \ln P}\vert_{S}$, with $S$ the specific entropy, and $\nabla_{rad}=\frac{3 \kappa P L}{16 \pi a c G m T^{4}}$. This simply states that if the gradient required to transport solely through radiation the energy going through a layer of stellar material is above the adiabatic gradient defined by the local thermodynamical properties of the stellar plasma, convection will take over and transport the energy outwards. While this criterion seems simple in essence, its implications and implementation are not so straightforward. First, the Schwarzschild criterion, as every other criterion available, is defined locally, while the properties of convective motions are intrisically non-local. Second, it is a dynamical condition, linked to the cancellation of the acceleration of the convective elements, defined originally from considerations within the so-called mixing-length theory. This implies that no form of convective penetration is taken into account. Third, the implementation of the convective boundary criteria was found to be erroneous in some stellar evolution codes and in contradiction with the authors who derived the said criterion \citep[see][for a discussion]{Gabriel2014}. 

If one wishes to take into account the stabilizing effects of mean molecular weight gradients or rotation (assuming cylindrical symmetry), the criterion is supplemented by additional terms, which define the so-called Ledoux and Soldberg-Hoiland criteria for both chemical gradients and rotation respectively. These criteria are written
\begin{eqnarray}
\nabla_{\mathrm{rad}}>\nabla_{\mathrm{ad}}+\frac{\phi}{\delta}\nabla_{\mu}, \\
\nabla_{\mathrm{rad}}>\nabla_{\mathrm{ad}}+\frac{\phi}{\delta}\nabla_{\mu}+\frac{H_{P}}{g \delta}\frac{1}{\varpi^{3}}\frac{d \left( \Omega^{2} \varpi^{4} \right)}{d\varpi}\sin \vartheta,
\end{eqnarray}
with $\delta=-\frac{\partial \ln \rho}{\partial \ln T}\vert_{P,\mu}$, $\phi=\frac{\partial \ln \rho}{\partial \ln \mu}\vert_{P,T}$, $\nabla_{\mu}=\frac{d \ln \mu}{d \ln P}\vert_{\rho,T}$, $g$ the local gravitational acceleration, $H_{P}$ the local pressure scale height, $\varpi$ the distance to the rotation axis, $\Omega$ the rotational velocity and $\vartheta$ the colatitude. 

Another approach to derive the border of convective regions has also been proposed in \citet{Roxburgh1992} and can be seen as providing an upper limit for the border position. 

\subsection{Formalism of convection}

The impossibility to directly model hydrodynamical phenomena in stellar models leads to the use of approximations and parametric approaches. In the case of convection, various formalisms have been developed, the most commonly used being the so-called mixing-length theory (MLT). This approach can be dated back to Ludwig Prandtl and stellar evolution codes usually refer to the works of \citet{Biermann1932,Biermann1948}, \citet{Bohm} or \citet{Cox} for the methodology they use. While widely used, the MLT is a very crude approximation of convective motion, where only radial displacements of the fluid element are considered, with all fluid elements having the same characteristic size which is supposed to be much smaller than any other length scale in the star, pressure equilibrium is maintained throughout the motion and the thermodynamic properties of the fluid element only differ slightly from its surroundings. In the upper layers of the stellar convective envelopes, this hypothesis does not hold and non-adiabatic effects have to be included in the formalism.

With all these hypotheses, the MLT considers that the fluid elements propagate over a certain mean free path before dissipating, the so-called mixing length. This mixing length defines one of the key convection parameter in stellar evolution, denoted $\alpha_{\mathrm{MLT}}$, which impacts the efficiency of the convective mixing. There are actually many other parameters within the MLT formalism, defining the so-called MLT ``flavour'' \citep[see e.g.][for a brief discussion]{Salaris}, but the most commonly known and tweaked is $\alpha_{\mathrm{MLT}}$. This parameter was defined as a constant, to be empiricaly calibrated using solar constraints. These so-called ``solar-calibrated'' $\alpha_{\mathrm{MLT}}$\footnote{A value of $\alpha_{\mathrm{MLT}}$ provided by calibrating a $1M_{\odot}$ model of a given initial chemical composition to reproduce the solar luminosity, radius and surface composition at the solar age.} values were then used to compute large grids of stellar models spanning all the regions of the HR diagram. With the advent of hydrodynamical simulations, this picture changed and various studies provided $\alpha_{\mathrm{MLT}}$ values calibrated from the simulations for various positions in the HR diagram \citep{Trampedach,Magic2015,Sonoi2019}. While these studies provide a way to add some flexibility and consistency to the MLT, they do not address the issue of the intrinsic shortcomings of this theory. 

\begin{figure*}
	\centering
		\includegraphics[width=12cm]{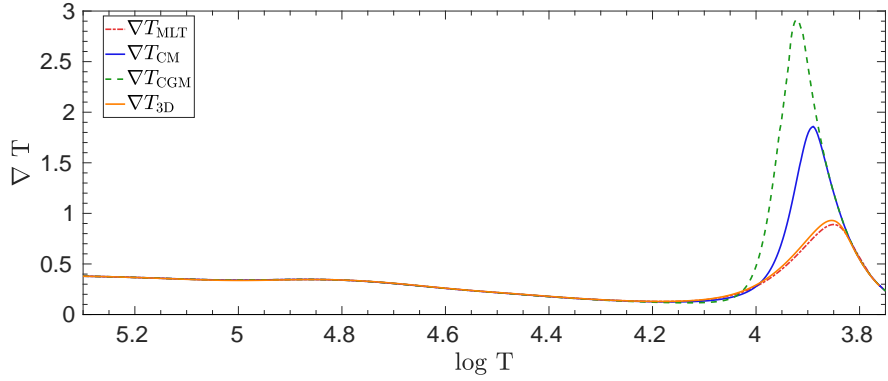}
	\caption{Temperature gradient profiles as of function of $r/R$ for solar models using different formalisms of convection, MLT stands for the classical mixing length theory, CM for the formalism of \citet{Canuto91}, CGM for the formalism of \citet{Canuto96} and $3D$ for a patched solar model from \citet{Sonoi2015}. }
		\label{FigNablaT}
\end{figure*}

Besides the ``classical'' MLT, other approaches have been developed by \citep{Canuto91, Canuto96}. This formalism considers that a spectrum of convective elements of various size, moving on a length scale comparable to the distance to the closest convective boundary. Both approaches are compared in terms of their temperature gradient to hydrodynamical simulations, as illustrated in figure \ref{FigNablaT}. As can be seen, the MLT model is far closer to the 3D hydrodynamical simulation than the so-called FST approach. This result must however be mitigated by the fact that the agreement in entropy gradient is far better for the FST model than for the MLT model \citep{Sonoi2019}. Ultimately, none of the formalisms agree perfectly neither with hydrodynamical simulations nor with helioseismic investigations. 

More sophisticated models of convections, using Reynold's stresses, have also been developed \citep{Xiong1985,Xiong1986}. This approach separates the convective flows into a mean contribution to which perturbations are added. While this approach certainly adds a degree of accuracy to the description of convection and has found to improve the agreement with solar frequencies, it also increases greatly the number of free parameters and equations associated with convection. Hence, it has never been included in stellar evolution codes and tested on large scales. 

\subsection{Convective boundaries and overshooting}

As previously discussed, the current description of convection in stellar interiors defines locally convective boundaries by the point where the acceleration of the elements is zero while the real boundary would be the point where their velocity is zero. This defines to the so-called ``overshooting'' issue, which is related to quantifying the amount and the form of extra-mixing that should be included at the border of convective regions described by the MLT. 

Various descriptions of the phenomenon are available. The most commonly used prescription is to add a small fully mixed region beyond the Schwarzschild limit derived from the MLT. The temperature gradient in this region is then set to be adiabatic or radiative. In the former case, one considers that the thermalisation of the convective elements that penetrate beyond the Schwarzschild limit will enforce an adiabatic stratification, while in the latter case, one assumes that the transport of energy is carried out by radiation only.  Another popular approach is to use the diffusive overshoot formalism devised by \citet{Freytag}. However, it should be noted that this approach does not describe overshooting in the classical sense, that is, the penetration of convective elements beyond the Schwarzschild limit, but rather the shear induced diffusive-like mixing that is generated from the turning of convective elements. Other prescriptions, based on hydrodynamical simulations, have been proposed by \citep{Meakin2007,Arnett2009,Viallet2013} . These ``turbulent entrainement'' prescriptions have been tested in various specific cases \citep[see e.g.][]{Cristini2016,Cristini2019} but are yet to be implemented in various evolutionary codes to test their impact on evolutionary models. 

Additional approaches such as those of \citet{Rempel2004}, \citet{Xiong2001}, \citet{Zhang2014}, \citet{Gabriel2018} have also been proposed but they have not been tested in full evolutionary calculations but only in specific cases, as for example in an analysis of the base of the solar convective zone in \citet{JCD2011}.

The main issue with overshooting is that it can significantly affect results from stellar modelling at various evolutionary stages. Indeed, it will affect the amount of fuel available for the nuclear reactions and thus extent evolutionary stages during which the star presents a convective core. It will thus impact the age of the star, its mass-luminosity ratio as well as the products of nucleosynthesis... The intensity of the required mixing can be constrained in two ways, using either eclipsing binaries or seismology. In the presence of gravity modes or mixed modes, seismic studies can even constrain the sharpness of the transition, as it will affect the oscillation cavity of gravity modes and hence their period spacing, as first demonstrated by \citet{Miglio2008}. Studies by \citet{Montalban2013, Bossini2015,Bossini2017} have also shown the potential of seismology to constrain extra-mixing in helium burning phases.

\subsection{Semi-convection}

In addition to overshooting, another form of extra-mixing, called semi-convection, can occur at the border of convective zones. This issue was already identified in the early days of stellar modelling \citep{Schwarzschild1958} but we are still lacking today a definitive answer on this issue. The origin of the problem is bound to the convective boundary criteria. For instance, in the case of convective core growing in mass during the main sequence, a hydrogen discontinuity builds up at the convective boundary and a small region above a convective core can be found to be convectively unstable under the Schwarzschild criterion but stable under the Ledoux criterion. In massive stars, the near equality between the adiabatic and radiative temperature gradient in the layers with a mean molecular weight gradient left by the receding convective core can induce a succession of semi-convective zones.

This leads to some confusion in the way to treat the mixing of chemical elements in this region. \citet{Kato1966} found that it should be the seat of overstable oscillations, which would lead to the mixing of chemical elements. When confronted with the issue, \citet{Schwarzschild1958}, \citet{Sakashita1961} and \citet{Castellani1971b} suggested to implement an efficient mixing of the elements until convective neutrality is reached, in other words, until $\nabla_{\mathrm{Ad}}=\nabla_{\mathrm{Rad}}$. \citet{Langer1983} suggested to treat the mixing using an additional diffusion coefficients. Other studies by \citet{Spruit1992}, \citet{Spruit2013A} and \citet{Zaussinger2013} found the mixing to be inefficient and suggested that it would actually be dominated by the mixing induced by rotation, whereas \citet{Wood2013} found the mixing to be very efficient, with transitory structured features in the semi-convective region. 

From an evolutionary point of view, adding mixing in the semi-convective region is similar to considering a form of overshooting, it will mainly extend evolutionary timescales for low-mass and massive  main-sequence and core-helium burning stars. From a seismic point of view, semi-convection will affect the propagation cavities of gravity modes and affect their period spacing. Hence, there is a strong potentially of seismology to constrain the intensity of this mixing. Moreover, the recent advances in seismology of white dwarfs may help us constrain semi-convection by revealing the traces of this mixing on the structure of these dying stars \citep{Giammichele2017, Giammichele2018}.  

\section{Rotation}

The second most uncertain transport mechanism acting in stars is rotation. Indeed, we know that stars do rotate (some of them very fast) and that this will affect their evolution, chemical abundances, stability and seismic properties. Rotation acts as a transport process of both angular momentum and chemical elements. The first main effect of rotational mixing is to act as an inhibitor of microscopic diffusion. Consequently, models including rotation show abundance variations with respect to their non-rotating counterparts (see  \citet{Brott2011}, \citet{Brott2011b},\citet{Ekstrom2012}). The effect of rotation has been extensively studied and the interested reader is referred to the book by \citet{Maeder2009} or the series of paper \citep{Maeder2012,Meynet2017} for further information. In the following section, I will briefly scratch the surface of this issue. 

\subsection{Formalisms and implementation}

First, it should be noted that the implementation of rotation in stellar models has geometrical considerations. The assumed spherical symmetry of stellar models is strictly speaking broken by the effects of rotation. However, while 2D models are currently under development \citep{EspinosaLara2013, Rieutord2016}, most of the evolutionary codes including the effects of rotation are 1D and consider the average deviations induced by rotation on the properties of stellar models. It is also worth mentioning that rotation is always implemented under the hypothesis of shellular rotation, meaning that both chemical composition and rotational velocity are constant along isobars, due to the very strong anysotropy of turbulent transport, resulting from the stable stratification of the stellar structure in radiative zones. This leads to changes in the structure equations in terms of both mechanical and thermal equilibrium.

However, the most complicated issue with rotating stellar models is the implementation of the transport of angular momentum and chemicals induced by rotation. This implies the introduction of an additional equation for the transport of angular momentum as well as an additional term in the equation of the evolution of chemical abundances. 

Generally, rotating 1D models include two main effects, that of the meridional circulation, resulting from the thermal imbalance between the equator and the poles, and that of shear induced turbulence, which serves as a regulator of the circulation. Two approaches to model these processes in radiative zones can be found in the litterature, the so-called ``advective-diffusive'' approach and the ``diffusive'' approach. The former one considers the transport of angular momentum to be governed by the combination of an advective transport by the circulation, following the formalism of \citet{Zahn1992}, and a diffusive transport by shear while the latter one considers a purely diffusive transport, following \citet{Endal1978,Endal1979,Endal1981}. In both approaches, the transport of chemicals induced by rotation is considered to be diffusive. 

Hence, the equations describing angular momentum transport and chemical elements transport under the effects of rotation are

\begin{eqnarray}
\frac{\partial \Omega}{\partial t}=\frac{1}{\rho r^{4}}\frac{\partial}{\partial r}\left( \rho r^{4} \nu \frac{\partial \Omega}{\partial r} \right), \\
\frac{\partial c_{i}}{\partial t}= \frac{1}{\rho r^{2}}\frac{\partial}{\partial r}\left(\rho r^{2} D \frac{\partial c_{i}}{\partial r}\right),
\end{eqnarray}

\noindent using the diffusive formalism. With $\Omega$ the rotational velocity, $\rho$ the local density, $r$ the radial position, $c_{i}$ the mass fraction of a given chemical element and $\nu$ and $D$ are the total viscosity and diffusion coefficient for the processes considered. In the advective-diffusive approach and assuming shellular rotation, one has 

\begin{eqnarray}
\frac{d r^{2}\Omega}{dt}=\frac{1}{5\rho r^{2}}\frac{\partial}{\partial r} \left( r^{4}\rho \Omega U \right)+\frac{1}{\rho r^{4}}\frac{\partial}{\partial r}\left( \rho r^{4} \nu \frac{\partial \Omega}{\partial r} \right),\\
\frac{dc_{i}}{dt}=\frac{1}{\rho r^{2}}\frac{\partial}{\partial r} \left( r^{2}\rho \left( D_{\mathrm{eff}}+D_{\mathrm{V}}\right) \frac{\partial c}{\partial r}\right),
\end{eqnarray}

\noindent with $\Omega$, $c_{i}$, $\rho$ and $r$ denoting the same quantities as before, $U$ is the advective velocity of the circulation currents, $\nu$ the diffusion coefficient of angular momentum and $D_{\mathrm{eff}}$ and $D_{\mathrm{V}}$ denotes the diffusion coefficients of chemicals resulting from vertical shear and efficient horizontal turbulence and the circulation. It has been indeed shown by \citet{Chaboyer1992} that the transport of chemical elements by rotation could indeed be treated as a diffusive process in radiative zones. This is a consequence of the fact that, unlike the transport of angular momentum, the transport of chemicals does not depend on the distance to the rotation axis of the star. 

In convective regions, the transport of angular momentum is much more complicated to describe, as these regions are crudely described in stellar evolution code. Helioseismic constraints \citep{Kosovichev1988} have shown that the Sun presents a rotation profile in the convective envelope that varies mostly with latitude and not so much in the radial direction. In general, solid-body rotation is used to describe these regions but simulations have shown that very deep envelopes in red giants could undergo differential rotation.

The existence of a self-consistent approach to model rotational transport in stars is a great success of modern stellar physics. However, the issue with the current formalism is that it doesn't work when confronted to various observations such as the solar rotation profile, the lithium abundances in stellar clusters and the seismic inferences on rotation in subgiant and red-giant stars. I will further discuss these issues in Section \ref{secRotaUncertainties}.

\subsection{Effects on structure and evolution}

The impact of rotation will differ with the type of star, the initial rotation velocities and initial metallicity of the models considered. General trends can be drawn, but the precise characteristics of the models will of course depend on the transport processes considered along the evolution. 

As a general trend, including rotation will lead to cooler and less-luminous models on the ZAMS, as shown in \citet{Eggenberger2012}. During the main sequence, the rotational-induced mixing will provide more hydrogen to the core, thus extending the duration of this part of the life of the star. In the case of low mass stars, for example for a typical $1M_{\odot}$ model, the evolutionary tracks on the main sequence will be nearly identical but as soon as a convective core is present during the evolution, the differences become much more significant. From \citet{Eggenberger2010a}, we can see that the impact of rotational mixing on the main sequence tracks is nearly the same as adding overshooting. The issue in disentangling both effects is very difficult as lifting the degeneracy would require to be able to isolate the different fine changes in the structure of the models. In addition, the reproduction of the lithium abundances in stellar clusters requires the inclusion of breaking by magnetized winds \citep{CantoMartins2011} or internal gravity waves \citep{Charbonnel1999}. In their study, \citet{Vick2010} also discuss the potential degeneracy in some cases between the effects of a turbulent mixing (potentially induced by rotation) and mass loss to reproduce the peculiar abundance patterns of AmFm stars.

Seismic indices are also impacted by rotational mixing, as shown in \citet{Eggenberger2010a, Eggenberger2010c}, frequency ratios and separations can be significantly affected as a result of the impact of rotational mixing on the stellar structure. Similarly, the evolution as a function of time of the surface helium abundance of a rotating solar model is significantly flatter than that of a non-rotating model, due to the inhibition of microscopic diffusion by rotation. However, the tests presented in \citet{Eggenberger2010a, Eggenberger2010c}, where these large deviations are observed, are for rotating models only including the effects of circulation and shear-induced turbulence, which exhibit a strong differential rotation in the radiative envelope. As we know from helio- and asteroseismic studies, these models are in contradiction with observations. 

If the impact of magnetic instabilities, required to reproduce the solar rotation profile, is taken into account, the differences in both helium abundance and helioseismic indices between rotating and non-rotating models become much smaller. The discrepancies are of course maximal at the end of the main-sequence, but they remain smaller than observational uncertainties. This behaviour is a result of the inhibition of shear by the Tayler-Spruit dynamo. It can be expected that internal gravity waves, which also flatten the internal rotation profile, will lead to a similar behaviour of the seismic indices. 

Rotational mixing is also expected to alter the period spacing of g-mode pulsators by altering the chemical stratification near the border of a convective core. Moreover, it has been shown by \citet{Bouabid2013} that under the effects of rotation, the period spacing of $\gamma$ Doradus stars would not oscillate around a constant value, but that its slope for the prograde, zonal and retrograde mode varies with the rotation velocity. Hence, as shown by \citep{Ouazzani2017,Ouazzani2018}, the slope of the period spacing can be used as direct diagnostic of internal rotation on the main-sequence, providing a testbed for models of angular momentum transport. 

For massive stars, rotation will also impact the occurence of mass-loss episodes and their intensity, which will change the abundance patterns at the surface (typically He, C, N and O). However, even without mass loss, the surface abundances of rotating models are significantly different from non-rotating ones, as can be seen from the comparisons performed in \citet{Ekstrom2012}. For massive stars, the abundance pattern of rotating models is in better agreement than for their non-rotating counterpart. However, the VLT FLAMES survey of massive stars has unraveled a behaviour in the nitrogen enrichment of B stars that still challenges the current rotating evolutionary models of single stars \citep{Hunter2008}. The issue is related to the fact that there is exist a significant fraction of slowly rotating stars exhibiting a large enrichment in nitrogen as well as a group of stars rotating fast but showing little to no chemical enrichment. These behaviours are unexpected and the reproduction of the so-called ``Hunter diagram'' is still a challenge for stellar population syntheses of single stars. The main difficulty is that the properties of rotating models are very similar to those exhibited by close binary systems and that many uncertainties still reside as to the impact and types of interactions in such system. It seems however clear that, due to the large fraction of massive stars in binary systems ($\approx 0.5$ according to \citet{Sana2008,Sana2009,Bosch2009,Sana2011}) that this should play a role. However, calculations by \citet{Meynet2011} have shown that magnetic braking could also lead to a slow rotation at the surface while simultaneously inducing a strong differential rotation in the interior, generating the observed chemical enrichment. In their paper, \citet{Brott2011b} also mention the possibility that the mechanism of \citet{MacGregor2003} could be responsible for the different behaviours observed in the survey.

In the HR diagram, blue loops crossing the Cepheid instability strip for intermediate mass stars will also occur at a different luminosity (higher or lower depending on the competing effects of the mixing and the centrifugal force) and the maximal luminosity of red supergiant stars is twice as low as that of non-rotating models.

\subsection{Additional processes and uncertainties} \label{secRotaUncertainties}

As discussed before, the current state of the modelling of rotation in stellar models has shown significant disagreements when confronted to observations. One of the most well-known issue is the inability of models to reproduce the solid-body rotating deep radiative layers of the Sun. Indeed, they predict a very strong differential rotation gradient in the radiative layers. To agree with helioseismic constraints, the transport of angular momentum must include either the effect of gravity waves \citep{Charbonnel2005} or the Tayler-Spruit magnetic instability \citep{Spruit1999,Spruit2002,Eggenberger2005}. Currently, both effects allow to reconcile models with helioseismology and the inferences from pressure modes do not allow to favour one solution over the other. The observations of solar gravity modes, if confirmed independently, will however strongly influence the outcome of this long-lasting issue.

With the advent of space-based photometry mission, the rapid development of asteroseismology also exposed strong disagreements in the rotating models of post-main sequence stars. Indeed, asteroseismic constraints have shown that these stars rotate up to a factor 80 slower than what is obtained from current rotating stellar models \citep{Deheuvels2012,Mosser2012,Deheuvels2014,Deheuvels2015,Gehan2018}. Various additional processes have been suggested to reconcile the models with seismic observations. Amongst them, plume-induced gravity waves have been shown to produce good results by \citet{Pincon2017} but more sophisticated tests based on their inclusion in stellar evolution codes is required to prove their efficiency. Azimuthal magneto-rotational instabilities have also been suggested as a potential candidate by \citet{Spada2016}. Besides these studies, investigations by \citet{Eggenberger2017,Eggenberger2019} have been carried out to constrain the efficiency of the missing process(es). 

Besides these well-known issues to seismologists, rotating models must also be supplemented by the braking by magnetized winds to reproduce the lithium abundances of stellar clusters. Gravity waves have also been proposed to reproduce the observed so-called lithium gap. 

From a theoretical perspective, various instabilities could also be at play \citep[see the book by][for an extensive review]{Maeder2009}. However, while additional processes are suggested to reconcile models and observations, refinements to purely hydrodynamical processes are also crucial to fully quantify the intensity of new ingredients \citep[see][]{Prat2014}. In addition, besides the contribution of potential additional processes, it is also their interactions and couplings that have to be properly modelled in stellar evolution codes. 

\section{Mass Loss}

Mass loss is a key element of stellar evolution. For low and intermediate mass stars, it is negligible for most of their lives and, with the exception of transfer in multiple systems, becomes only significant on the upper parts of the RGB and AGB. For massive stars, around $~20M_{\odot}$\footnote{The exact number varies with the stellar metallicity, the figure of $20M_{\odot}$ being for a solar metallicity value.}, the situation is very different as they are subject to substantial line-driven wind that leads to a significant peeling of their upper layers already during the main-sequence. Hence, for stars above $40M_{\odot}$ at solar metallicity, most of the stellar lifetime will occur on the blue side of the HR diagram, and the stars will never reach the red supergiant stage\footnote{Strong internal mixing can also lead to homogeneous evolution, which leads to a similar behaviour.}. 

The main difficulty with the modelling of mass loss is the variety of phenomena that can be at play. For example, the wind of main-sequence massive stars is driven by UV lines, whereas dust will be the main driver in red supergiant stars. These processes will of course be affected by both rotation and pulsations, which can enhance the mass loss rate of a given star. In addition, eruptive events can also lead to dramatic shedding of the star's upper layers in a very short time, which is very different from the steady implementation used in stellar evolution codes.

In recent years, the paradigm on stellar mass loss has somewhat changed, due to the discovery of an overestimation of observational mass loss rates due to the so-called ``clumping'' effect, which I will discuss later, the increased importance of eruptive events and the high fraction of massive binary stars. It is of course beyond the scope of this brief review to discuss all of these discoveries thoroughly, hence I refer the interested reader to the reviews by \citet{Iben1983}, \citet{Chiosi1986}, \citet{Langer2012} and \citet{Smith2014} and references therein for a far more detailed discussion and additional references. 

\subsection{Issues and implementation}

As presented above, mass loss can have various origins and thus different behaviours depending on the physical mechanism driving the wind. 

One of the recent thorny issue of mass loss in stellar models is the presence of clumps, small scale inhomogeneities which lead to biases in the empirical formulas used in evolutionary codes. When fitting the laws to the observational constraints, most of the formulas assume an homogeneous wind, which is not expected from a theoretical point of view, and not observed in practice. Physically, the presence of clumps in the wind implies that these small regions will scatter light more efficiently than if the matter was uniformly distributed. Therefore, when assuming an homogeneous wind, one overestimates the density of the wind to reproduce the high emissions observed due to the clumps. 

This effect is expected from a theoretical point as a result of line-driven instability or by sub-surface convection resulting from the iron opacity bump. Large-scale inhomogeneities can also be appear due to fast rotation, time-variations of absorbing component or the influence of the stellar magnetic fields on the repartition of the matter in the wind. 

Various studies have been dedicated to quantify a correction, denoted as a ``clumping'' factor, that is used to amend the current empirical laws that overestimate the observed mass loss rates \citep[see e.g.][]{Bouret2005, Fullerton2006, Puls2006}.

In classical stellar evolution code, mass loss is implemented as a steady rate, depending on mass, temperature, luminosity, metallicity and sometimes the terminal velocity of the wind. These expressions are mostly empirical and, as can be seen from figure 2 of \citet{Beasor2018}, produce significantly different trends. Classical approaches include the empirical law of \citet{Reimers1975}:

\begin{eqnarray}
\dot{M}_{\mathrm{R}}=-4 \times 10^{-13}\eta \frac{L}{gR}\frac{g_{\odot}R_{\odot}}{L_{\odot}}
\end{eqnarray}

which was calibrated from the observations of red giants with approximately solar metallicity. There is no solid theoretical background for Reimers' formula and it is mostly useful to draw order-of-magnitude estimates. This formula has been refined over the years to take into account more detailed dependencies in stellar parameters as is that of \citet{Schroder2005}:

\begin{eqnarray}
\dot{M}_{\mathrm{S-C}}=-8\times 10^{-14}\frac{LR_{\odot}M_{\odot}}{L_{\odot}RM}\left(\frac{T_{\mathrm{eff}}}{4000} \right)^{3.5} \left(1+\frac{g}{4.3 g_{\odot}} \right)
\end{eqnarray}

Both these formulas are dedicated to reproduce the mass loss of red giants. As stated before, the empirical laws will be adjusted depending on the stars considered and their expected regimes. For massive stars, one can note the formulations of, amongst others, \citet{Nieuwenhuijzen1990}, \citet{Bloecker1995}, \citet{Lamers1981}, \citet{Jager1988}, \citet{Nugis2000} or \citet{Vink2001}.

\subsection{Impact on stellar models}

In this section, I will briefly discuss the impact of mass-loss at various evolutionary stages. I remain qualitative in our description, as the details will vary with the exact stellar parameters and the parametrization chosen for the mass-loss rate. General trends can however be outlined. I separate our discussion between low-mass stars and high-mass stars, where the distinction is mainly made by the importance of mass loss on the main sequence. 

\subsubsection{Intermediate and low-mass stars}

Mass loss has been mentioned as a potential sources of the ``anomalies'' observed in some pre-main sequence and main-sequence stars \citep{Vick2011}, if the selective effects of the radiative acceleration on the elements are considered. This leads to an issue in constraining the actual origin of the so-called ``chemical peculiarities'' in AmFm stars or the Spite plateau \citep{Spite1982,Vick2010,Vick2013}, as they can be expressed by a combination of mass-loss, rotation, and selection effects of microscopic diffusion. Constraints on mass loss on the RGB has been obtained by \citet{Handberg2017} using asteroseismic constraints for the open cluster NGC 6819. They studied the efficiency of the mechanism on the RGB and found a total mass lost of $0.03\pm0.01$ M$_{\odot}$ on the RGB by looking at the mean masses of cluster members on the RGB and in the clump. Mass loss has also been invoked to explain the presence of so-called ``helium-rich'' stars \citep{Vauclair1975} but this explanation has been refuted over the years \citep{Krticka2006}.

On the AGB, mass loss will control the maximum luminosity on the branch, lifetimes and chemical yields. Thus, it plays a central role in population syntheses and understanding the properties of stellar clusters, for example the ratio of stars on the horizontal branch and on the AGB). Recent results by \citet{Cassisi2016} have shown that, contrary to what was previously believed, a highly efficient mass-loss was not required to reproduce this ratio. However, a moderate mass loss is required on the RGB to reproduce the HB morphology in globular clusters \citep{Salaris2016}. This is of course in contradiction with the results by \citet{Handberg2017} which found no significant mass loss on the RGB in NGC 6819. On the horizontal branch, mass loss has been invoked as a potential formation channel of SdB stars \citep{Yong2000}. However, \citet{Vink2002} have shown that this scenario was invalidated but could explain the distribution of rotation velocities and element abundances observed in SdB and HB stars.

Mass loss is of course important during the planetary nebula phase, but also for white dwarfs. In the latter case, the effects of radiative accelerations would induce the apparition of a weak wind and would modify the abundances of metals in the atmosphere \citep{Vennes2006,Wilson2019} although accretion of circumstellar material are also suspected to provide fresh elements at the surface of the white dwarf.

\subsubsection{Massive stars}

As mentioned before, mass-loss will significantly affect global parameters. However, it will also leave clear marks in the surface abundances. First, for stars undergoing combustion  of hydrogen through the CNO cycle, the processed material will become apparent at the surface, as a result of the peeling of the envelope. If this takes place for a sufficiently long time, products of helium fusion may even appear at the surface. On the main-sequence, the effects of mass-loss in massive stars will be to decrease the central temperature, resulting from the reduction of the central pressure. The evolution of the convective core will thus be directly affected: its size and that of the semi-convective region at its top will be smaller. 

These two effects impact the duration of the main-sequence in an opposite way. The lower central temperature will reduce the luminosity and the efficiency of nuclear burning whereas the reduced size of the mixed region of chemicals will reduce the amount of fuel available. Unless large mass-loss rates are used, the reduction of the central temperature has a larger impact and the main-sequence lifetime is increased. However, the size of the helium core at the end of the main-sequence is smaller than in stars evolving at constant mass.

While the luminosity at a given time of models including mass-loss is lower than that they would have if they evolved at constant mass, it is however higher than that of a model having the mass they have at this specific point in time. Thus, isochrones and lines of constant mass are changed. The higher the masses and ages of the models, the higher the biases mass-loss can induce on determinations from evolutionary tracks. The stars undergoing the strongest mass loss rates will remain in the blue side of the HR diagram, not passing through the red supergiant phase and procude WR stars. Typically, stars with masses below $20M_{\odot}$ will not exhibit sufficient mass loss and go through a red supergiant phase before their explosion. At the transition between these extremes, one can find intermediate cases where the stars undergo a supergiant phase before evoling back to the blue side and even becoming a WR star before the end of their nuclear burning time. 

The behaviour of models during the helium burning phase under the effects of mass loss is slightly more complicated. Various competing effects can be at play and will strongly influence the global properties of the more massive models. Mass loss is, for example, one of the parameters that come into play when explaining the crossing of the so-called Hertzsprung gap. Of course, the behaviour of the track will be strongly dependent on the assumed mass-loss rates and thus for example the ratio of the helium burning phase over the hydrogren burning phase may also largely vary with these assumptions. For more massive stars, the behaviour of mass loss during the helium burning phase is also very important to understand the different populations of Wolf-Rayet stars and the progenitors of supernovae events (see for example \citet{Meynet2015} and \citet{Beasor2018} for both a theoretical perspective and a re-determination of the observed mass loss rates of RSG).  

In later stages, mass loss will become less and less important for the massive stars, as the evolutionary timescales become shorter. Hence, in a first approximation, they can be considered as evolving at constant mass. 

On a general note, one key issue for understanding the properties of mass loss is to quantify its dependency with stellar metallicity. This factor is crucial to understand the origin of the differences between various stellar populations formed in different environments. While the impact of the metallicity on the intensity of line-driven winds is clear, it also play a crucial role in the formation of dust in extended envelopes and through its impact on opacity, on the occurrence of Roche Lobe Overflows in close binary systems. 

\section{Conclusion}

In this brief review, I have discussed some of the key physical processes currently contributing to the uncertainties of stellar models. It is obvious that some of the issues we face are long-lasting and that they result from the interaction of complex phenomena. Their understanding will require further theoretical work, while their implementation in stellar evolution codes will rely on innovative solutions to reproduce the diversity and couplings that can be at play.

Today, stellar physics is blessed with datasets of unprecented quality. From a modelling perspective, the challenge is to use these data in the most optimal way by improving inference techniques. With better inferences, we may provide meaningful constraints on the models. Simultaneously, theoretical advances will provide physically relevant parametrization of the currently uncertain mechanisms. In the current era of astrophysics, it is of utmost importance to state that a higher accuracy on stellar parameters will only stem from theoretical improvements of stellar models. 

\section*{Acknowledgments}
G.B. acknowledges support from the ERC Consolidator Grant funding scheme ({\em project ASTEROCHRONOMETRY}, G.A. n. 772293). This work is sponsored the Swiss National Science Foundation (project number 200020-172505).

\bibliographystyle{phostproc}
\bibliography{Buldgen.bib}

\end{document}